\documentclass[twocolumn,showpacs,preprintnumbers,amsmath,amssymb]{revtex4}

\usepackage{graphicx}
\usepackage{dcolumn}
\usepackage{bm}

\usepackage{amssymb}

\newcommand{\ee}{\end{equation}} 
\newcommand{\be}{\begin{equation}} 
\def\spose#1{\hbox to 0pt{#1\hss}}
\def\ltapprox{\mathrel{\spose{\lower 3pt\hbox{$\mathchar"218$}}
 \raise 2.0pt\hbox{$\mathchar"13C$}}}
\def\gtapprox{\mathrel{\spose{\lower 3pt\hbox{$\mathchar"218$}}
 \raise 2.0pt\hbox{$\mathchar"13E$}}}

\begin{document}


\title{Constraints on the IR behavior of the ghost propagator
        in Yang-Mills theories}

\author{A. Cucchieri}
\affiliation{Instituto de F\'{\i}sica de S\~ao Carlos, Universidade de
             S\~ao Paulo, \\ Caixa Postal 369, 13560-970 S\~ao Carlos, SP, Brazil}
\author{T. Mendes\footnote{Permanent address: 
             Instituto de F\'{\i}sica de S\~ao Carlos, Universidade de
             S\~ao  Paulo, C.P.\ 369, 13560-970 S\~ao Carlos, SP, Brazil.}
        }
\affiliation{DESY--Zeuthen, Platanenallee 6, 15738 Zeuthen, Germany}

\date{\today}

\begin{abstract}
We present rigorous upper and lower bounds for the momentum-space
ghost propagator $G(p)$ of Yang-Mills theories in terms of the smallest
nonzero eigenvalue (and of the corresponding eigenvector) of the Faddeev-Popov
matrix. We apply our analysis to data from simulations of SU(2) lattice 
gauge theory in Landau gauge, using the largest lattice sizes to date.
Our results suggest that, in three and in four space-time
dimensions, the Landau-gauge ghost propagator is {\em not} enhanced
as compared to its tree-level behavior. This is also seen in plots and fits
of the ghost dressing function. In the two-dimensional case, on the other hand,
we find that $G(p)$ diverges as $p^{-2-2\kappa}$ with $\kappa
\approx 0.15 \, $, in agreement with Ref.\ \cite{Maas:2007uv}. We note
that our discussion is general, although we make an application only
to pure gauge theory in Landau gauge. Most of our simulations have
been performed on the IBM supercomputer at the University of S\~ao Paulo.
\end{abstract}

\maketitle


\section{Introduction}
\label{sec:intro}

In the Gribov-Zwanziger scenario \cite{Gribov},
confinement of quarks (in Landau and in Coulomb gauge) is related
to a ghost propagator $G(p)$ enhanced in the infrared (IR) limit
when compared to the tree-level propagator $1/p^2$. An enhanced ghost
propagator is also required by the Kugo-Ojima criterion \cite{Kugo}
in order to explain confinement of color charge (in Landau gauge).

In order to investigate the origin of this IR enhancement
we consider, following Ref.\
\cite{Cucchieri:2006hi}, a generic gauge condition ${\cal F}[A]=0$,
where $A$ is the gauge field. The condition is imposed on the lattice
by minimizing a functional $E[U]$, where $U$ is the (gauge)
link variable. From the second variation
of $E[U]$ one can define the Faddeev-Popov (FP) matrix ${\cal M}(a,x;b,y)$.
(Here we use $a$ and $b$ to indicate color indices, whereas $x$ and $y$
represent points of the lattice.)
In the SU($N_c$) case, ${\cal M}$ is an $(N_c^2 - 1)V \times
(N_c^2 - 1)V$ matrix, where $V=N^d$ is the lattice volume, given
by the number of points $N$ along each side of the lattice and by
the space-time dimension $d$. This matrix is real and symmetric
with respect to the exchange $(a, x) \leftrightarrow (b, y)$ (see
Ref.\ \cite{Cucchieri:2005yr} for a thorough discussion of the properties
of the FP matrix in Landau gauge). At any (local) minimum of
$E[U]$, all the eigenvalues of ${\cal M}$ are positive (modulo
trivial null eigenvalues). The set of all minima of $E[U]$ is the
so-called Gribov region $\Omega$.
On the boundary of $\Omega$ ---  known as the first 
Gribov horizon $\partial \Omega$ --- the
smallest nontrivial eigenvalue $\lambda_{min}$ of the FP matrix
is null. Since the configuration space has very large
dimensionality we expect that, in the limit of large lattice volumes $V$,
entropy favors configurations near $\partial \Omega$ \cite{Greensite:2004ur,
horizon,Cucchieri:1997ns}, i.e.\ $\lambda_{min}$ should go to
zero in this limit. This is indeed the case in 2d \cite{Maas:2007uv},
3d \cite{Cucchieri:2006tf} and 4d Landau gauge \cite{Sternbeck:2005vs},
in 4d Coulomb gauge \cite{Greensite:2004ur} and in 4d Maximally Abelian
gauge (MAG) \cite{Mendes:2006kc}.

The ghost propagator is written in terms of the inverse of the FP matrix
as
\begin{equation}
 G(p)\, = \,\frac{1}{N_c^2 - 1} \sum_{x\mbox{,}\, y\mbox{,}\, a}
                 \frac{e^{- 2 \pi i \, k \cdot (x - y)}}{V}\,
        {\cal M}^{- 1}(a,x;a,y) \, ,
\label{eq:Gp}
\end{equation}
where the lattice momentum $p(k)$ has components $p_{\mu}(k)
= 2 \sin{(\pi k_{\mu} /N)}$ with
$k_{\mu} = 0, 1, \ldots, N-1$. Also, here and in the formulae below,
a path-integral average is understood.
Since ${\cal M}(a,x;b,y)$ develops a null eigenvalue at
the Gribov horizon $\,\partial \Omega \,$, one might expect that
the corresponding ghost propagator
diverges at small momenta in the infinite-volume limit. This
could in turn introduce a long-range effect in the theory, 
related to the color-confinement mechanism \cite{Greensite:2004ur,
horizon}. Indeed, in
several numerical studies (using relatively small lattice volumes),
the ghost propagator $G(p)$ seems to be IR-enhanced in 3d
\cite{Cucchieri:2006tf} and 4d Landau gauge \cite{Cucchieri:1997dx,ghost}
and in 4d Coulomb gauge \cite{Langfeld:2004qs}.
On the other hand, in MAG \cite{Mendes:2006kc}
one finds an IR-finite $G(p)$. At the same time, recent results in
Landau gauge using very large lattice volumes \cite{Cucchieri:2007md,big,Sternbeck:2007ug}
suggest an essentially tree-level-like IR ghost propagator $G(p)$ in 3d and in 4d.
(A similar result has been obtained in Refs.\ \cite{analytic,Boucaud} using
different analytic approaches.)
Finally, in the 2d case, one finds \cite{Maas:2007uv} $G(p) \sim
p^{-2.3}$ after extrapolating data for the IR exponent $\kappa$
to infinite volume. Therefore, we have cases for which $G(p)$ is not IR-enhanced 
in the infinite-volume limit and thus the argument reported above
cannot be valid in general \cite{Cucchieri:2006hi}. The aim of this work
is to understand under what conditions one should expect an IR-enhanced ghost
propagator $G(p)$. This is done by investigating constraints on the behavior
of $G(p)$ and thus obtaining better control over its infinite-volume
extrapolation.
Clearly, since the only diverging quantity (besides the lattice volume)
is $1/\lambda_{min}$, the IR enhancement of $G(p)$ may be
closely related to this eigenvalue and to the projection of its corresponding
eigenvector $\, \psi_{min}(a,x) \,$ on the plane waves $ e^{- 2 \pi i \, k \cdot x}/
\sqrt{V}$ for small momenta $p(k)$.


\section{Lower and upper bounds for $G(p)$}
\label{sec:ineq}

In this section we obtain upper and lower bounds for the momentum-space
ghost propagator $G(p)$ in terms of the smallest nonzero eigenvalue
$\lambda_{min}$ [and of the corresponding eigenvector $\, \psi_{min}(a,x) \,$]
of the FP matrix ${\cal M}(a,x;b,y)$. To this end, let us
first introduce our notation. As usual, we define the generic eigenvalue
$\lambda_i$ of ${\cal M}(a,x;b,y)$ and the corresponding eigenvector
$\psi_i(a,x)$ by using the relation
$ {\cal M}(a,x;b,y) \, \psi_i(b,y) = \lambda_i \, \psi_i(a,x) $,
where the index $i$ takes values $1, 2, \ldots, (N_c^2 - 1)V$
and the sum over repeated indices is understood.
As stressed in the Introduction, this matrix is positive
(or semi-positive) definite, whenever we use a minimizing condition in order
to fix a gauge (minimal gauge). Also, since it is obtained from a second-order
expansion, this matrix can always be written in a symmetric form. Then, the
eigenvectors $\psi_i(a,x)$ can be assumed orthogonal to each other and
normalized as $ \sum_{a, x} \, \psi_i(a,x) \, \psi_j(a,x)^* = \delta_{i j}$,
where $^*$ indicates complex conjugation.
Using this notation we can write
\begin{equation}
{\cal M}(a,x;b,y) \; = \sum_{i, \lambda_i \neq 0}
   \lambda_i \, \psi_i(a,x) \, \psi_i(b,y)^* \; .
\end{equation}
Note that we are working in the space orthogonal to the kernel of
the FP matrix. Then we can write the inverse of this matrix as \cite{barton}
\begin{equation}
{\cal M}^{-1}(a,x;b,y) \; = \sum_{i, \lambda_i \neq 0}
   \lambda_i^{-1} \, \psi_i(a,x) \, \psi_i(b,y)^* \; .
\end{equation}
By using Eq.\ (\ref{eq:Gp}) we obtain
\begin{equation}
G(p) \; = \; \frac{1}{N_c^2 - 1} \, \sum_{i, \lambda_i \neq 0} \,
                      \sum_a \, \lambda_i^{-1} \,
                    | {\widetilde \psi_i(a,p)} |^2 \; ,
\end{equation}
where ${\widetilde \psi_i(a,p)} = \sum_x
\psi_i(a,x) \, e^{- 2 \pi i k \cdot x} / \sqrt{V}$
is the Fourier transform of the eigenvector $\psi_i(a,x)$.
Since all the nonzero eigenvalues are positive and $ 0 < \lambda_{min}
\leq \lambda_i $, we can write the inequalities
\begin{eqnarray}
& & \,\,\,\,\,\, \frac{1}{N_c^2 - 1} \, \frac{1}{\lambda_{min}} \, \sum_a \,
  | {\widetilde \psi_{min}(a,p)} |^2
   \, \leq \, G(p) \; , \\[1mm]
& & G(p) \, \leq \, \frac{1}{N_c^2 - 1} \, \frac{1}{\lambda_{min}} \,
              \sum_{i, \lambda_i \neq 0} \, \sum_a \,
                 | {\widetilde \psi_i(a,p)} \, |^2 \; .
\label{eq:ineq2ini}
\end{eqnarray}
(Note that we assume nondegenerate eigenvalues. However, the second 
inequality applies also when $\lambda_{min}$ is degenerate and the first 
one can be easily modified to take a degeneracy into account.)
By summing and subtracting in Eq.\ (\ref{eq:ineq2ini})
the contributions from the eigenvectors corresponding
to a null eigenvalue and using the completeness relation
$ \sum_i \, \psi_i(a,x) \, \psi_i(b,y)^*
= \delta_{ab} \delta_{xy}$, we find
\begin{equation}
G(p) \, \leq \, \frac{1}{\lambda_{min}} \, \left[ \, 1
                  \,-\, \frac{1}{N_c^2 - 1} \sum_{j, \lambda_j = 0} \, \sum_a \,
                | {\widetilde \psi_j(a,p)} \, |^2 \, \right] \; .
\label{eq:ineq2}
\end{equation}
Let us stress that the above results are simply a consequence of the positivity of the
FP matrix, i.e.\ they apply to gauge-fixed configurations that belong to the
interior of the first Gribov region.


\begin{table}
\vspace{-3mm}
\begin{center}
\caption{\label{tab1}
The ghost propagator $G(p_s)$ for the smallest nonzero momentum $p_s = 2 \sin(\pi / N)$
and the inverse of the smallest nonzero eigenvalue $\lambda_{min}$ of the FP matrix
for various lattice volumes and $\beta$ values in the 3d case. Data (in physical
units) are taken from Ref.\ \cite{Cucchieri:2006tf}.}
\vspace{1mm}
\begin{tabular}{|c|c|c|c|c|}
\hline
$N$ & $\beta$ & $a^{-1}$ [GeV] & $G(p_s)$ [GeV$^{-2}] $ &
                                       $1/\lambda_{min}$ [GeV$^{-2}] $ \\
\hline
 20 &  6.0 & 1.733 &  7.3(1) & 32.5(7) \\
 30 &  6.0 & 1.733 & 19.4(3) & 94(2) \\
 20 &  4.2 & 1.136 & 21.0(4) & 107(3) \\
 30 &  4.2 & 1.136 & 54.5(8) & 282(6) \\
\hline
\end{tabular}
\end{center}
\vspace{-5mm}
\end{table}

\section{Bounds in Landau gauge}
\label{sec:landau}

In the Landau case the eigenvectors corresponding to the zero eigenvalue are constant modes,
i.e.\ they contribute only to the case $p = 0$.
Thus, for any nonzero momentum we have
\begin{equation}
\frac{1}{N_c^2 - 1} \, \frac{1}{\lambda_{min}} \, \sum_a \,
  | {\widetilde \psi_{min}(a,p)} |^2
   \, \leq \, G(p) \, \leq \, \frac{1}{\lambda_{min}} \; .
\label{eq:ineq}
\end{equation}
As said in the Introduction, in the infinite-volume limit the measure gets
concentrated on the boundary of the (first) Gribov region, 
i.e.\ $\lambda_{min}$ goes to zero in this limit. 
In particular, as for the Laplacian operator, one can expect to find
$\, \lambda_{min} \,\sim\, N^{-\alpha} \,$ for large $N$.
Now, if we make the hypothesis of a power-law behavior $\, p^{-2 - 2 \kappa} \,$
for $G(p)$ in the IR limit and consider
the smallest nonzero momentum on the lattice [i.e.\ $p_s = 2 \sin(\pi / N)$],
we have $G(p_s) \sim N^{2+2\kappa}$ in the limit of large $N$.
Then from Eq.\ (\ref{eq:ineq}) we get $2+2\kappa \leq \alpha$. Thus,
assuming a power-law behavior for $G(p)$, $\alpha > 2$ is a necessary condition
to obtain a ghost propagator $G(p)$ enhanced in the IR limit
compared to the tree-level behavior $1/p^2$.

\begin{table}
\vspace{-3mm}
\begin{center}
\caption{\label{tab2}
As in Table \ref{tab1} for the 4d case.
Data (in lattice units) are taken from Refs.\ \cite{Cucchieri:1997dx,
Cucchieri:1997ns} using the average called {\em first copy}. We find similar results
when considering data obtained using the average called {\em absolute minimum}.}
\vspace{1mm}
\begin{tabular}{|c|c|c|c|}
\hline
$N$ & $\beta$ & $G(p_s)$ & $1/\lambda_{min}$ \\
\hline
 8 & 0.8 & 8.94(8) & 91(3) \\
 12 & 0.8 & 22.1(6) & 210(20) \\
 16 & 0.8 & 40.6(7) & 360(60) \\
 8 & 1.6 & 7.2(1) & 61(3) \\
 16 & 1.6 & 32.1(3) & 220(20) \\
 8 & 2.7 & 3.4(1) & 11.0(4) \\
 12 & 2.7 & 7.1(3) & 25(1) \\
 16 & 2.7 & 12.9(6) & 45(4) \\
\hline
\end{tabular}
\end{center}
\vspace{-5mm}
\end{table}

The lower bound in Eq.\ (\ref{eq:ineq}) clearly depends on the behavior of the quantity
$ | {\widetilde \psi_{min}(a,p_s)} |^2 $.
Note that this quantity cannot diverge in the infinite-volume
limit. (Actually, since the eigenvectors are normalized, this
quantity is always between 0 and 1.)
Thus, if we furthermore make the assumption of a behavior
at large $N$ given by $N^{-\gamma}$, with $\gamma \geq 0$, we find
the condition $\alpha - \gamma \leq 2+2\kappa$.

\begin{figure}
\vspace{-18mm}
\includegraphics[scale=0.38]{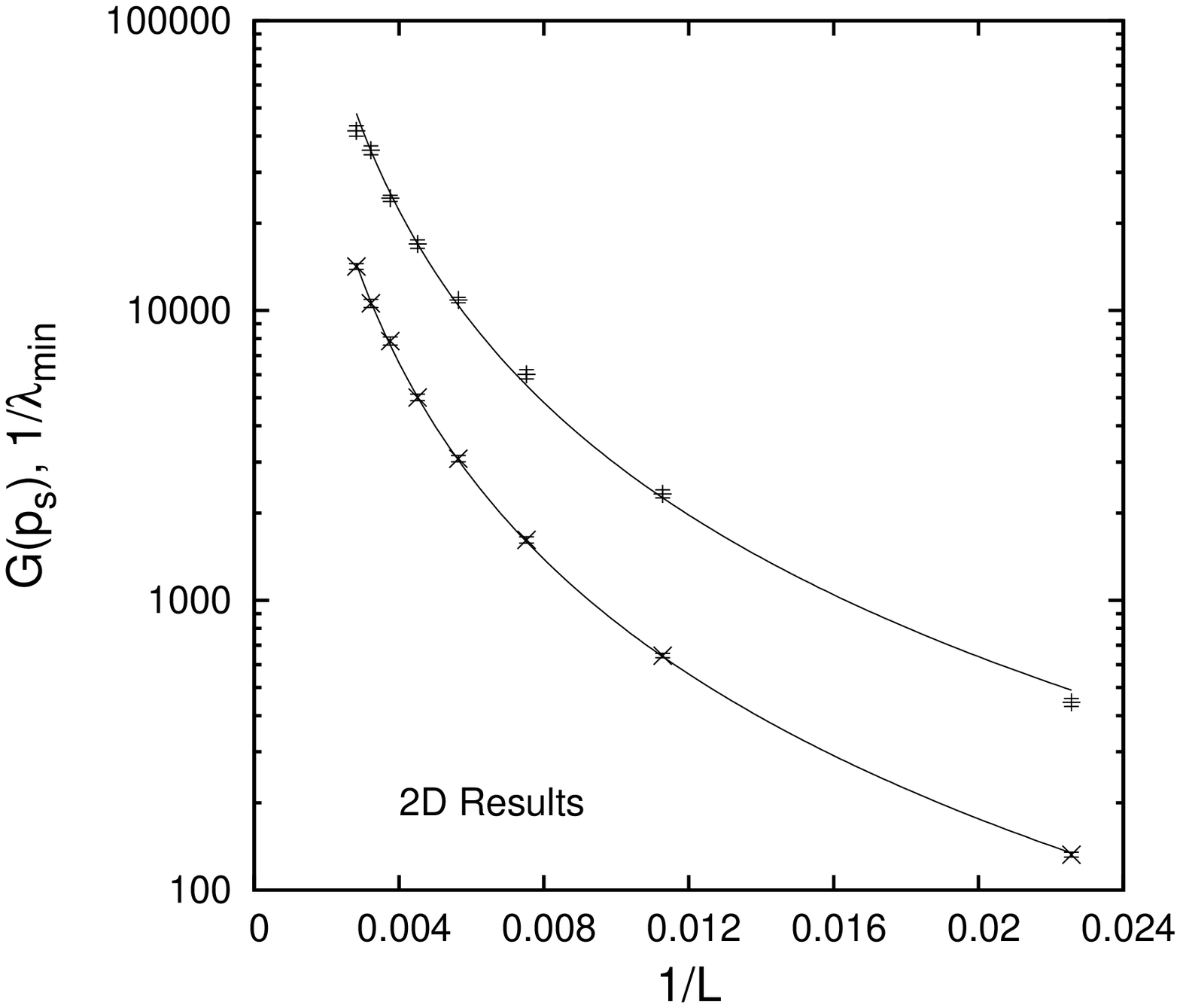}

\vspace{-35mm}
\hspace{1mm}
\includegraphics[scale=0.38]{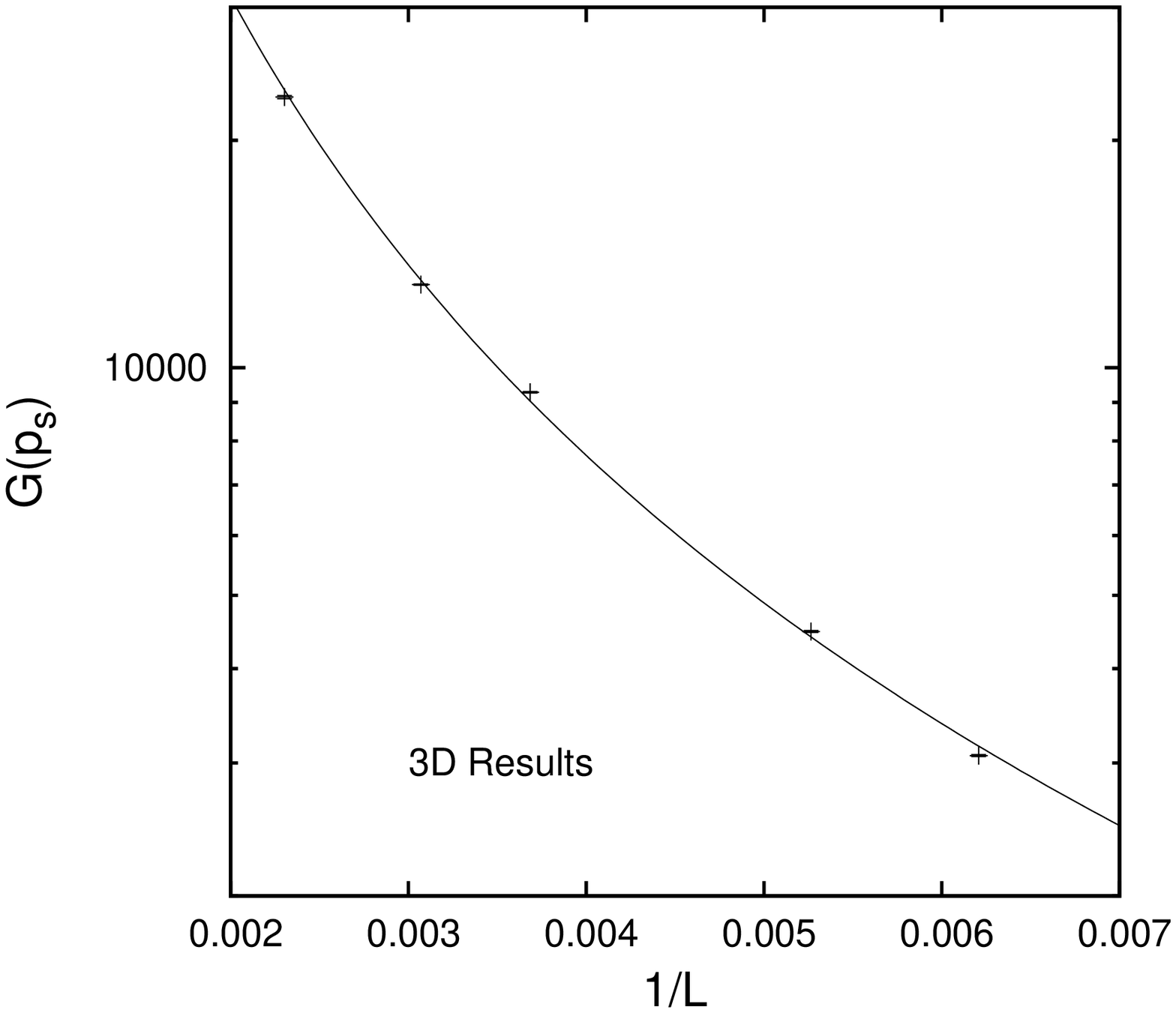}

\vspace{-35mm}
\hspace{1mm}
\includegraphics[scale=0.38]{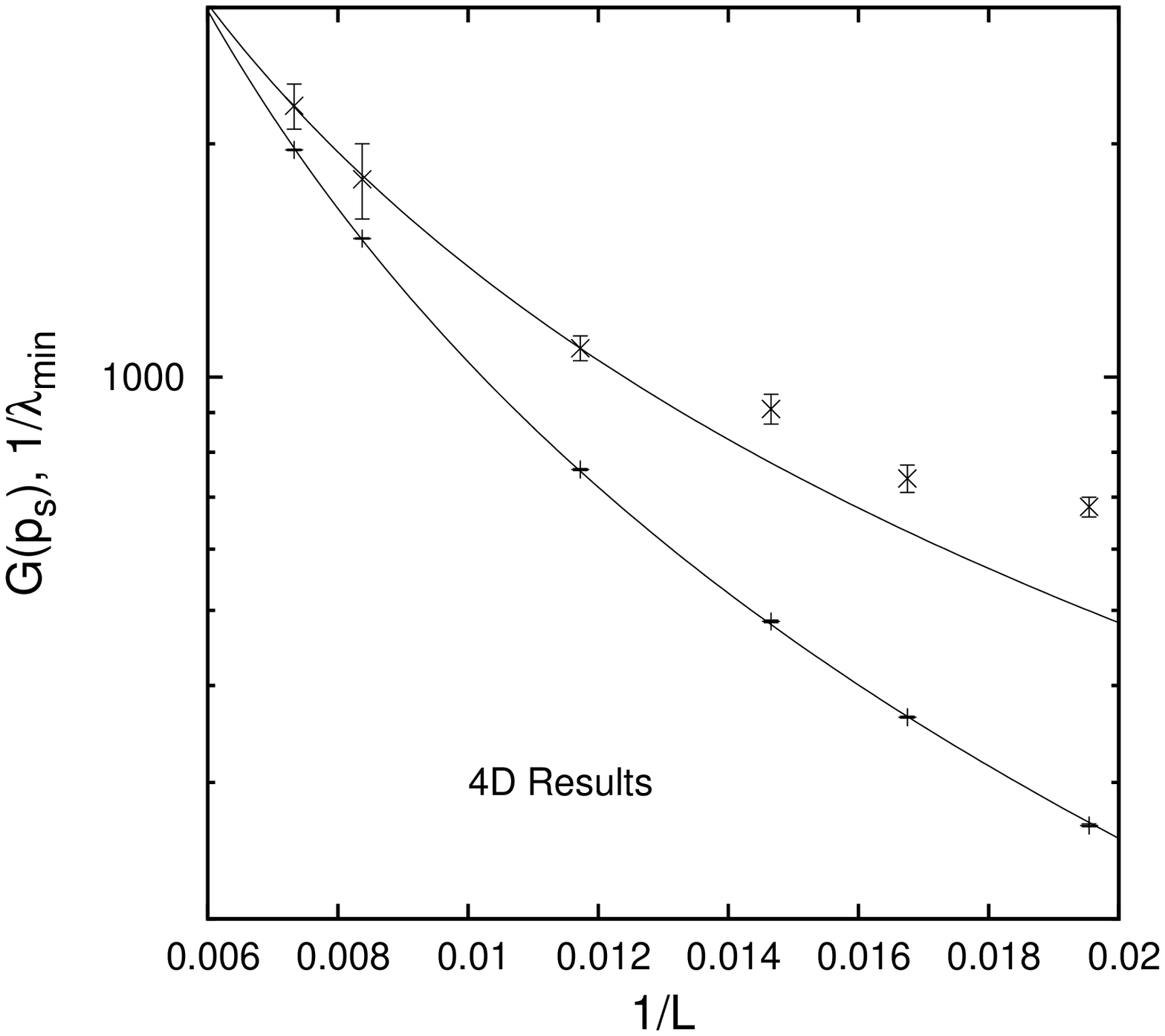}
\vspace{-15mm}
\caption{\label{fig:G}
  The ghost propagator $G(p_s)$ for the smallest nonzero momentum $p_s = 2 \sin(\pi / N)$
  (in GeV$^{-2}$) as a function of the inverse lattice side $1/L$ (GeV) for the
  2d case (top, at $\beta = 10$ with volumes up to $320^2$), 3d case (center,
  at $\beta = 3$, with volumes up to $320^3$) and the 4d case (bottom, at
  $\beta = 2.2$ with volumes up to $128^4$). Data are taken from
  Ref.\ \cite{Cucchieri:2007md} for the 3d and the 4d cases.
  We also show (in 2d and in 4d) the inverse of the smallest nonzero eigenvalue
  $\lambda_{min}$ of the FP matrix (in GeV$^{-2}$). In these two cases one can
  verify the inequality $1/\lambda_{min} \geq G(p_s)$. The fitting
  parameters are reported in Table \ref{tab:fits}. Note that we did not evaluate
  $\lambda_{min}$ for all our configurations.
  }
\end{figure}

As a check of the proposed bounds, we collect here 
results for the ghost propagator $G(p_s)$ at the smallest
nonzero momentum $p_s = 2 \sin(\pi / N)$ and for the smallest
nonzero eigenvalue $\lambda_{min}$ from Refs.\ 
\cite{Cucchieri:1997dx,Cucchieri:1997ns, Cucchieri:2006tf,Cucchieri:2007md} 
for the 3d and 4d cases. These data are shown for the smaller lattices
in Tables \ref{tab1}, \ref{tab2} and for the larger lattices in
Fig.\ \ref{fig:G}. We also present new
data in the two-dimensional case. (These data  were obtained together
with the data for the gluon propagator reported in \cite{Cucchieri:2007rg}.) 
All data refer to the $SU(2)$ case. Let us recall that recent studies
\cite{Sternbeck:2007ug,comparison}
have verified the analytic prediction that Landau-gauge gluon and ghost
propagators in SU(2) and in SU(3) are rather similar. Thus, we expect that
the analysis presented here should apply also to the SU(3) case.
In all cases the quantity $1/\lambda_{min}$ has been evaluated using the
average value for $\lambda_{min}$ and propagation of errors.

\begin{table}
\begin{center}
\caption{Fits of $G(p_s)$ and of $1/\lambda_{min}$
         respectively using the Ans\"atze
         $b \, L^{2+2\kappa}$ and $c \, L^{\alpha}$. 
         In the 4d case the fit for $1/\lambda_{min}$
         has been done by considering only the three largest
         physical volumes. When considering the three smallest
         physical volumes we find $\alpha = 0.9(3)$.
         In the other cases all data have been considered
         for the fit.
\label{tab:fits}}
\vspace{1mm}
\begin{tabular}{|c|c|c|c|c|}
\hline
 $d$ & $b$ & $2+2\kappa$ & $c$ & $\alpha$ \\
\hline
  2  & 0.026(1) & 2.251(9) & 0.12(3) & 2.20(4) \\
  3  & 0.11(3)  & 2.02(4)  &         &         \\
  4  & 0.086(3) & 2.043(8) & 1.2(1)  & 1.53(2) \\
\hline
\end{tabular}
\end{center}
\vspace{-5mm}
\end{table}

As one can see, the
upper bound in Eq.\ (\ref{eq:ineq}) is always satisfied. (Of course, this is
the case also when $p \neq p_s$.) The bound has also been verified using the data
at $\beta = 0$ from Refs.\ \cite{Cucchieri:1997dx,Cucchieri:1997ns}. We have fitted
the data for $G(p_s)$ and for $\lambda_{min}$ as a function of $L$ for the data
shown in Fig.\ \ref{fig:G} using a power-law Ansatz. It is interesting to note
that (in 2d and in 4d) we actually find $\alpha$ smaller than $2+2\kappa$. 
The bound is still satisfied, since the multiplicative constant in the fit of
$\lambda_{min}$ is larger. It would be interesting to see how the upper bound
is realized when even larger lattice volumes are considered. Also, note that 
the upper bound in the 4d case seems to saturate as the volume increases.
This would indicate that the contribution to $G(p)$ from all the eigenvalues
$\lambda_i > \lambda_{min}$ stays finite at large volume and that the exponent
$\gamma$, defined above, is zero. Let us note that this is not the case at
smaller lattice volumes, since then one needs to consider the contribution
to $G(p)$ from the first 150-200 smaller eigenvalues in order to
reproduce the behavior of the propagator at small momenta \cite{Sternbeck:2005vs}.
Finally, let us recall that finite-size effects for $G(p)$ (at fixed $p$)
are generally very small. This can be explained considering the spectral
density of the FP matrix, as done in Ref.\ \cite{Cucchieri:2006hi}.
Thus, our results indicate that in 3d and in 4d the ghost propagator is likely not
IR-enhanced, while in 2d we obtain $\kappa \approx 0.1$.
We note that our result in the 2d case is essentially in agreement with 
Ref.\ \cite{Maas:2007uv},
where however the value of $\kappa$ is obtained by first fitting $G(p)$
in an appropriate momentum range for each lattice volume and then
performing the extrapolation to infinite volume by a four-parameter
fit. The fits for $\lambda_{min}$ and $G(p_s)$ as a function of $L$ 
are usually simpler, as shown above.

\begin{figure}
\vspace{-17mm}
\includegraphics[scale=0.38]{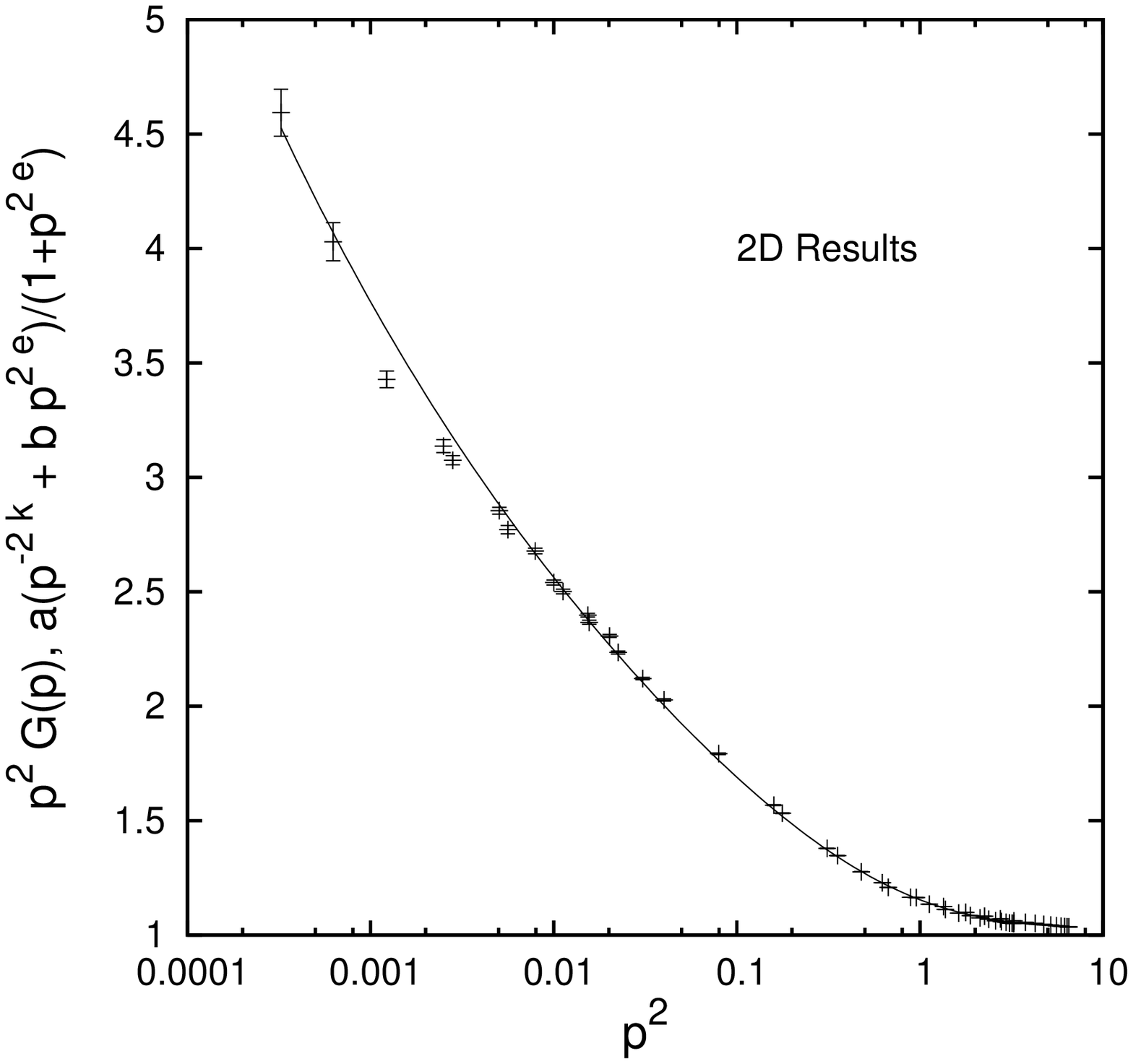}

\vspace{-34mm}
\hspace{1mm}
\includegraphics[scale=0.38]{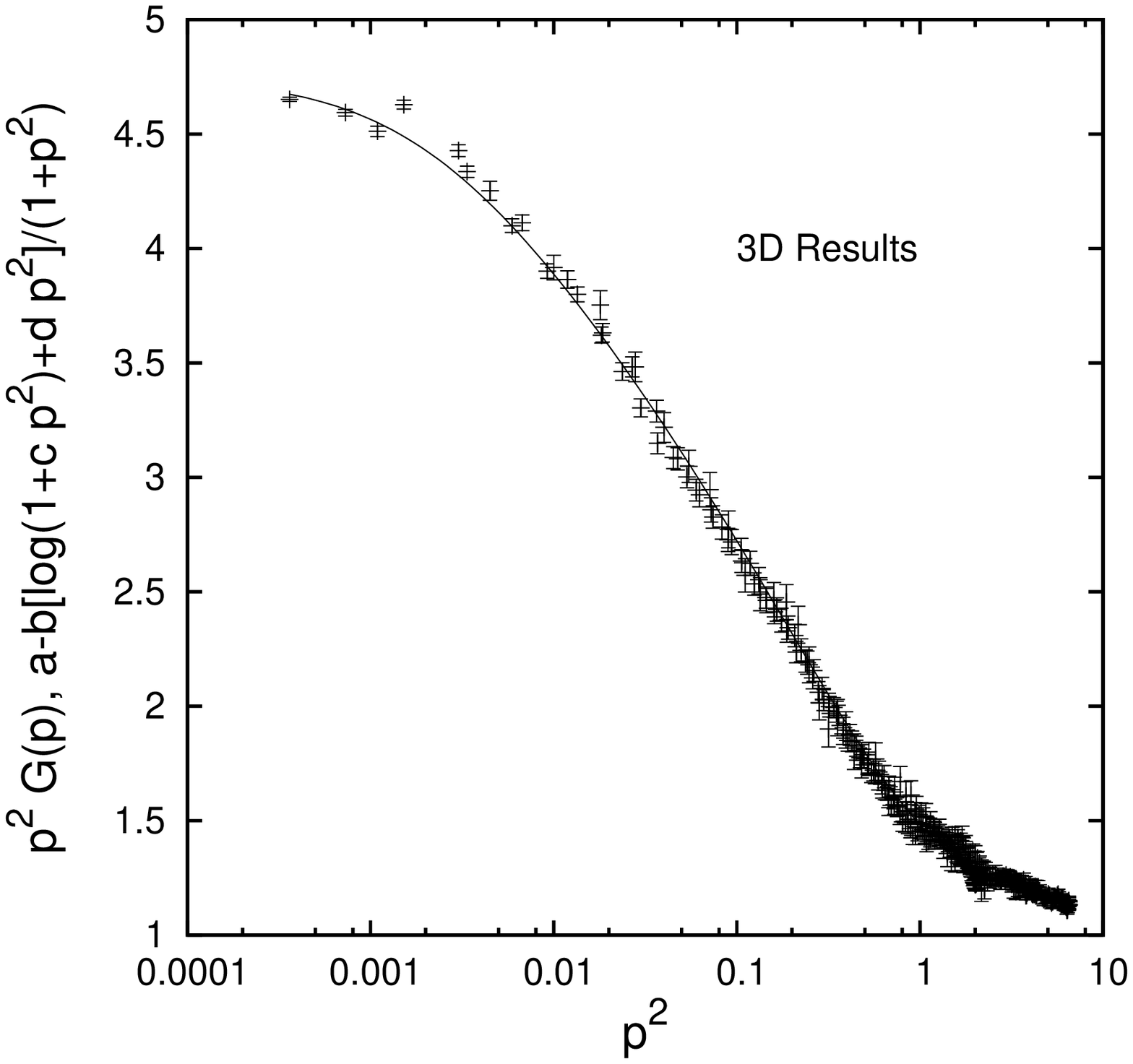}

\vspace{-34mm}
\hspace{1mm}
\includegraphics[scale=0.38]{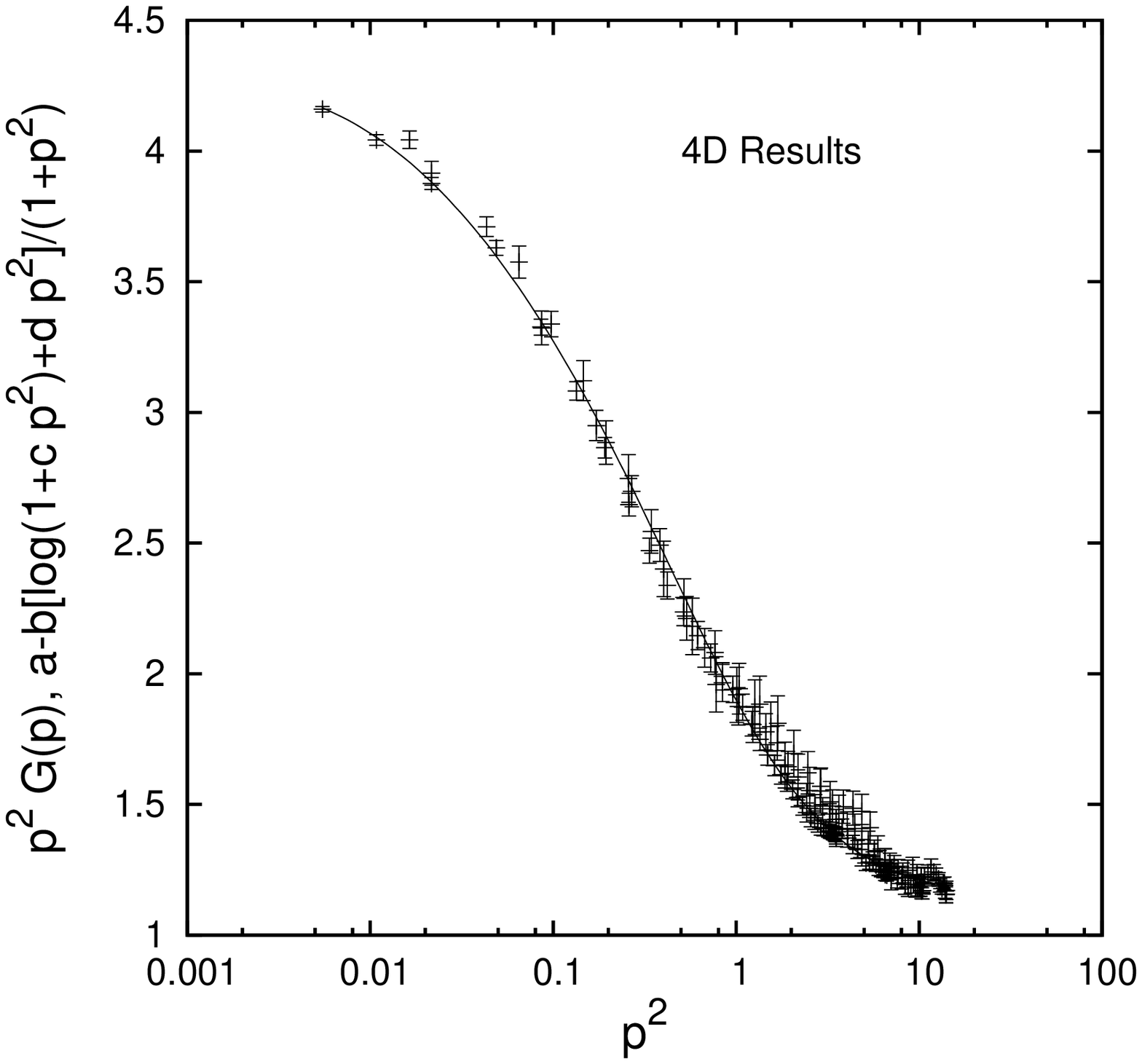}
\vspace{-19mm}
\caption{\label{fig:Gp2}
  The ghost dressing function $p^2 G(p^2)$ as a function of $p^2$
  (in GeV) for the
  2d case (top, at $\beta = 10$ with volume $320^2$), 3d case (center,
  at $\beta = 3$, with volume $240^3$) and the 4d case (bottom, at
  $\beta = 2.2$ with volume $80^4$). The fitting functions and the corresponding
  fitting parameters are reported in Table \ref{tab:fitsgp2bis}.
  }
\end{figure}

These results are confirmed if one considers the dressing function
$p^2 G(p^2)$ at the largest lattice volumes for the three cases 
(see Fig.\ \ref{fig:Gp2}). Indeed,
the data in the 2d case can be fitted using a power-law Ansatz
$a p^{-2 \kappa}$ (see Table \ref{tab:fitsgp2}),
with $\kappa = 0.177(2)$, essentially in agreement with \cite{Maas:2007uv}.
The fitted value for $\kappa$ decreases if one considers fewer points,
dropping points with larger $p$, and becomes $0.152(7)$ when 
considering data with $p^2 \in [0,0.1]$.
The same Ansatz does not work in 3d and in 4d. In these cases
the data for the dressing function can be described by the
Ansatz $a - b \log(1 + c p^2)$, inspired by the
logarithmic corrections suggested in Ref.\ \cite{Boucaud}. In both cases one finds
(approximately) $p^2 G(p^2) \to 4.5$ in the limit $p \to 0$.
This result supports the value $\kappa \approx 0$ obtained above,
when considering the bound using $\lambda_{min}$.
It is interesting that one can also obtain a relatively good fit
for the dressing function
in the whole range of momenta by considering the
fitting function $a(p^{-2k} + b p^{2e})/(1+p^{2e})$ in the 2d case
and $a - b [\log(1+c p^2) + d p^2] /(1+p^2)$ in 3d and in 4d. These fits
and the corresponding parameters are reported in Fig.\ \ref{fig:Gp2}
and in Table \ref{tab:fitsgp2bis}.

\begin{table}
\vspace{-3mm}
\begin{center}
\caption{Fits of $p^2 G(p^2)$ using the Ans\"atze
         $a p^{-2 \kappa}$ in the 2d case and
         $a - b \log(1 + c p^2)$ in the 3d and 4d cases.
         In the first case we used the data in the range $p^2 \in [0,0.5]$
         for the fit. In the other two cases the data in the range $p^2 \in [0,1]$
         have been considered.
\label{tab:fitsgp2}}
\vspace{1mm}
\begin{tabular}{|c|c|c|c|c|}
\hline
 $d$ & $a$ & $\kappa$ & $b$     & $c$ \\
\hline
  2  & 1.134(7) & 0.177(2) &          &         \\
  3  & 4.7(1)   &          & 0.579(5) & 320(20) \\
  4  & 4.28(1)  &          & 0.69(2)  & 33(3)   \\
\hline
\end{tabular}
\end{center}
\vspace{-5mm}
\end{table}

\begin{table}
\vspace{-4mm}
\begin{center}
\caption{Fits of $p^2 G(p^2)$ using the Ans\"atze
         $a(p^{-2k} + b p^{2e})/(1+p^{2e})$ in the 2d case and
         $a - b [\log(1+c p^2) + d p^2] /(1+p^2)$ in the 3d and 4d cases.
         In the three cases we used the whole range of momenta for the fit.
\label{tab:fitsgp2bis}}
\vspace{1mm}
\begin{tabular}{|c|c|c|c|c|c|c|}
\hline
 $d$ & $a$     & $\kappa$ & $b$      & $e$      & $c$     & $d$    \\
\hline
  2  & 1.24(3) & 0.16(2)  & 0.86(3)  & 0.75(15) &         &        \\
  3  & 4.75(1) &          & 0.491(5) &          & 450(30) & 7.1(1) \\
  4  & 4.32(2) &          & 0.38(1)  &          & 80(10)  & 8.2(3) \\
\hline
\end{tabular}
\end{center}
\vspace{-5mm}
\end{table}


\section{Conclusions}
\label{sec:con}

Our data suggest that the Landau-gauge ghost propagator does not
diverge faster than $1/p^2$ at small momenta in 3d and in 4d, while in 2d 
$G(p)$ does not diverge faster than $p^{-2-2\kappa}$ with $\kappa$
between 0.1 and 0.2,
in agreement with Ref.\ \cite{Maas:2007uv}. These results have been obtained by
considering the inequality in Eq.\ (\ref{eq:ineq2}) and by fitting the data
for $G(p_s)$ and $\lambda_{min}$ as a function of $L$. In particular,
we note that the use of the upper bound to constrain $G(p)$ is quite
convenient, since $\lambda_{min}$ does not depend on $p$. 
It might thus be of interest to optimize
the evaluation of $\lambda_{min}$ and of $\psi_{min}(a,x)$.

The results obtained here in the 3d and 4d cases,
together with those reported in Ref.\ \cite{Cucchieri:2007rg} for 
the gluon propagator, seem to contradict
the Gribov-Zwanziger confinement scenario for Landau gauge. On the other hand, 
one should establish if the behavior of gluon and ghost propagators at very small 
momenta is really so essential for the explanation of confinement. After all, 
when dynamical quarks are considered \footnote{Simulations in the
unquenched case \cite{ghost-unq} have shown that the behavior of
the ghost propagator is essentially not affected compared to the quenched case.}, string
breaking should manifest itself at a scale of about a fermi, i.e.\ for a energy scale
of about 200 MeV. Thus, for the physics of hadrons the behavior of the
propagators at intermediate momenta is probably more important. Let us recall that
for a space-time separation of about 1 fermi the transverse gluon propagator 
violates reflection positivity \cite{violation}, i.e.\ becomes
negative in real-space coordinates, and the exponent $\kappa$ 
for the ghost propagator is about 0.2-0.3 considering $p \sim 
0.5$ GeV \cite{Cucchieri:2006tf,Cucchieri:1997dx,ghost,Cucchieri:2007md,big,
Sternbeck:2007ug}. 
Thus, in the range 200-500 MeV, nonperturbative effects are clearly present in 
the behavior of the gluon and of the ghost propagator. 
For this range of momenta the use of our lower bound for the ghost
propagator could probably also be valuable.
Of course, the open problem is if these effects suffice to explain color confinement.


\section{Acknowledgements}

Discussions with several participants of the workshop on {\em Quarks and Hadrons
in strong QCD} (St.\ Goar) are acknowledged.
We also thank R.\ Sommer for helpful comments on the manuscript. 
We acknowledge partial support from FAPESP (under grant \# 05/59919-7) and
from CNPq (including grants \# 476221/2006-4 and 455353/2007-7). The work of T.M. is
supported also by a fellowship from the Alexander von Humboldt Foundation.
Most of the simulations reported here have been done on the IBM
supercomputer at S\~ao Paulo University (FAPESP grant \# 04/08928-3).



\end{document}